\documentclass[secthm,seceqn]{elsart}
\usepackage{amssymb}
\usepackage{amsmath}
\usepackage{graphicx}
\usepackage{mathrsfs} 

\newcommand{\R}{{\mathbb{R}}}

\newcommand{\remove}[1]{}
\newcommand{\god}{G\"odel}
\newcommand{\G}{{\mathscr{G}}}
\newcommand{\F}{{\mathscr{F}}}
\newcommand{\GL}{\mathbb{G}}
\newcommand{\Sub}{{\rm Sub}}

\newcommand{\Leaf}{{\rm Leaf}}
\newcommand{\Nover}{{\mathscr{T}}}

\newcommand{\Rus}{{\mathscr{R}}}
\newcommand{\rus}{{\mathscr{R}}}

\begin{document}

\begin{frontmatter}



\title{An Analysis of  Ruspini Partitions in G\"odel Logic}


\author[MAT]{Pietro Codara\corauthref{cor}},
\ead{codara@mat.unimi.it}
\author[DICO]{Ottavio  M. D'Antona},
\ead{dantona@dico.unimi.it}
\author[DICO]{Vincenzo Marra}
\ead{marra@dico.unimi.it}
\corauth[cor]{Corresponding author. Tel.: +39 02 50316312; Fax: +39 02 50316276.}
\address[MAT]{Dipartimento di Matematica F.~Enriques,
Universit\`a degli Studi di Milano,\\
via Saldini 50, I-20133 Milano, Italy}
\address[DICO]{Dipartimento di Informatica e Comunicazione,
Universit\`a degli Studi di Milano,\\
via Comelico 39, I-20135 Milano, Italy}

\begin{abstract}
By a \emph{Ruspini partition} we mean a finite family of fuzzy sets $\{f_1, \ldots, f_n\}$, $f_i : [0,1] \to [0,1]$,
such that $\sum_{i=1}^n f_i(x)=1$
for all $x \in [0,1]$, where $[0,1]$ denotes the real unit interval. We analyze such partitions in the language of
\god{} logic. Our first main result  identifies  the precise degree
 to which the Ruspini condition is expressible in this language, and yields {\it inter alia} a constructive procedure to
axiomatize a given Ruspini partition by a theory in \god{} logic. Our second main result extends this analysis to
Ruspini partitions fulfilling the natural additional condition that each $f_i$ has at most one left and one right
neighbour, meaning that $\min_{x \in [0,1]}{\{f_{i_1}(x),f_{i_2}(x),f_{i_3}(x)\}}=0$ holds for $i_1\neq i_2\neq i_3$.
\end{abstract}

\end{frontmatter}


%
\section{Introduction}\label{section:intro}

Let $[0,1]$ be the real unit interval.
By a \emph{fuzzy set} we shall mean a function $f \colon [0,1] \to [0,1]$.
Throughout the paper, we fix a finite nonempty family
\[
P=\{f_1, \ldots, f_n\}
\]
of fuzzy sets, for $n\geq 1$ an integer. Moreover, we write $\underline{n}$ for the set $\{1,\dots,n\}$.

 In several soft computing applications, the following notion of fuzzy partition plays an important role. It is
often traced back to \cite[p.\ 28]{ruspini}.

\begin{defn} We say
$P$ is a \emph{Ruspini partition} if  for all $x \in [0,1]$
\begin{equation}
\label{eq:Ruspini}
\sum_{i=1}^n{f_i(x)}=1\,.
\end{equation}
\end{defn}

\begin{figure}[ht!]
        \begin{center}
                \includegraphics{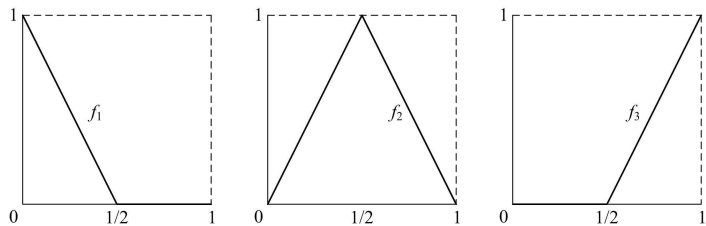}
        \end{center}
        \caption{A Ruspini partition $\{f_1,f_2,f_3\}$.}
        \label{fig:partition}
\end{figure}

\noindent  By way of informal motivation for what follows, think of the real unit
interval $[0,1]$  as the normalized range of values of a physical observable, say temperature.  Then
each $f_i \in P$ can be viewed as a means of assigning a truth-value to a
proposition about temperature in some many-valued logic $\mathscr{L}$.
Had one no information at all about such propositions, one would be led to identify them with propositional variables $X_i$,
subject only to the axioms of $\mathscr{L}$. However, the set $P$ does encode information about
$X_1, \ldots, X_n$.
For example, consider $P=\{f_1, f_2, f_3\}$ as in Fig. \ref{fig:partition},
and say $f_1$, $f_2$, and $f_3$ provide truth-values for
the propositions $X_1 = $ ``The temperature is low'', $X_2 = $ ``The temperature is medium'',
and $X_3 = $ ``The temperature is high'', respectively. If  $\mathscr{L}$ has a conjunction $\wedge$
interpreted by minimum, the proposition $X_1 \wedge X_3$ has $0$ as its only possible truth-value, \textit{i.e.},
it is a contradiction. The chosen set $P$ then leads one to add  extra-logical axioms
to $\mathscr{L}$, \textit{e.g.}, $\neg (X_1 \wedge X_3)$, in an attempt to express the
fact that one cannot observe both a high and a low temperature at the same time. More
generally, $P$ implicitly  encodes a \emph{theory}---that is, a family of formulas required
to hold, thought of as extra-logical axioms---over the pure logic $\mathscr{L}$. Imposing
the Ruspini condition on $P$, then, amounts to implicitly enriching the logic $\mathscr{L}$
by extra-logical axioms that
attempt to capture condition (\ref{eq:Ruspini}) in the language provided by $\mathscr{L}$.
Indeed, while  in practice it is often the case that $\mathscr{L}$ lacks the power to express
addition of real numbers exactly, $\mathscr{L}$ will still afford an approximation of the
Ruspini condition \emph{in its own language}. In this paper we are thus concerned with the
general problem of making explicit the extra-logical information implicitly encoded by $P$.

 Throughout this paper, we shall take $\mathscr{L}$ to be \god{} logic. Among triangular norms
 and conorms \cite{tnorms},  the minimum and maximum operators are rather popular  choices to
 model fuzzy logical conjunction and disjunction in applications. {\god} logic adds to this
 setting an implication that is obtained from conjunction via \emph{residuation}, and thus
 fits into P.\ H\'{a}jek's family of fuzzy logics based on (continuous) triangular norms;
 we refer to \cite{hajek}   for an extensive treatment.

 Here we recall that \emph{{\god} \textup{(}infinite-valued propositional\textup{)} logic} $\GL_\infty$
 can be syntactically defined as the schematic extension of the
intuitionistic  propositional calculus by the \emph{prelinearity axiom} $(\alpha \to \beta) \vee (\beta \to \alpha)$.
It can also be semantically defined as a many-valued  logic, as follows.
Let us consider well-formed formulas over propositional variables
$X_1,X_2,\dots$ in the language $\wedge,\vee,\to,\neg,\bot,\top$. (We use $\bot$ and $\top$
as the logical constants {\it falsum} and {\it verum}, respectively).
By an
\emph{assignment} we shall mean a function $\mu$
from (well-formed) formulas to $[0,1] \subseteq \R$
such that, for any two such formulas $\alpha,\ \beta$,

\begin{itemize}
        \item[] $\mu(\alpha \wedge \beta) = \min\{\mu(\alpha),\mu(\beta)\}$
        \item[] $\mu(\alpha \vee \beta) = \max\{\mu(\alpha),\mu(\beta)\}$
        \item[] $\mu(\alpha \rightarrow \beta) = \left\{\begin{array}{l}
        1\ \ \ \ \ \ \,\textrm{if}\ \mu(\alpha)\leq\mu(\beta)\\
        \mu(\beta)\ \ {\rm otherwise}
        \end{array}\right.$
\end{itemize}
and $\mu(\neg \alpha)=\mu(\alpha \to \bot)$, $\mu(\bot)=0$, $\mu(\top)=1$.
A \emph{tautology} is a formula $\alpha$ such that $\mu(\alpha)=1$ for every assignment $\mu$. As is well known, {\god} logic
is complete with respect to this many-valued semantics. Indeed, for $\alpha$ a formula of $\GL_\infty$, let
us write
$
 \vdash \alpha
$
to mean that $\alpha$ is derivable from the axioms of $\GL_\infty$ using \textit{modus ponens} as the only
deduction rule. Then the
completeness theorem guarantees that $\vdash \alpha$ holds if and only if $\alpha$ is a tautology.
For proofs and more details,  see  \cite{hajek}, \cite{gottwald}.

  This paper provides a thorough analysis of how the Ruspini condition on $P$ is
  reflected by its associated theory over \god{} logic. In (\ref{eq:F-alphaPnormalform})
  we shall eventually obtain a constructive procedure to
axiomatize the theory implicitly encoded by $P$.  \god{} logic cannot precisely capture
addition of real numbers, and Theorem \ref{th:main}  in fact proves that---up to logical
equivalence in $\GL_\infty$---the
Ruspini condition (\ref{eq:Ruspini})  reduces to the notion of \emph{weak Ruspini partition}
given in
Definition \ref{def:weak ruspini}. In Section \ref{sec:preliminary} we collect the necessary
algebraic and combinatorial
background, and prove some preliminary results.
Theorem \ref{th:main} is proved in Section \ref{s:main}.

\begin{figure}[ht!]
        \begin{center}
                \includegraphics{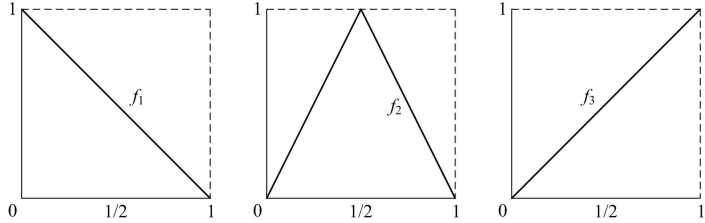}
        \end{center}
        \caption{A $3$-overlapping  family $\{f_1,f_2,f_3\}$.}
\label{fig:partition2}
\end{figure}

 In several applications, the family of fuzzy sets $P$ satisfies additional requirements beyond the Ruspini
condition.
Indeed, designers
often prefer fuzzy sets that  have at most one neighbour to the left and one neighbour to the right,
as in Figure \ref{fig:partition}. If, by contrast, one allows configurations
such as the one in Figure \ref{fig:partition2}, one
 contemplates the possibility that certain  values of the physical observable---temperature,
 in our example---are at
the same time low, medium, and  high (to possibly different degrees). While this may
be what is called for by specific situations, it turns out that in many applications the membership
functions are chosen so as to avoid this. Cf.\ \textit{e.g.}, the majority of the examples in \cite{yenlan}.
Formally, we consider the following definition.
\begin{defn}
 We say $P$ is \emph{$2$-overlapping} if for all $x \in [0,1]$ and all triples of indices $i_1\neq i_2\neq i_3$ one has
\begin{equation}\label{eq:2overlap}
\min{\{f_{i_1}(x),f_{i_2}(x),f_{i_3}(x)\}}=0 \ .
\end{equation}
\end{defn}

The set $P$ in Figure \ref{fig:partition}, for instance, is a $2$-overlapping family.
One could define \emph{$k$-overlapping} families of fuzzy sets in the obvious manner.
However, in this paper we shall only deal with the $2$-overlapping case.

 In Section \ref{s:mainbis}, we subject a family $P$ of $2$-overlapping fuzzy sets to the same
 analysis carried out for the Ruspini condition. Indeed, Theorem \ref{th:main2} is the exact
 counterpart for condition (\ref{eq:2overlap}) of Theorem \ref{th:main}. There are, however,
 two significant differences. Firstly, {\god} logic does capture the minimum of two real
 numbers exactly. This is why we do not need a weakened notion of $2$-overlapping families
 of fuzzy sets, whereas for the Ruspini condition  the concept of a weak Ruspini partition
given in Definition \ref{def:weak ruspini} is unavoidable. Secondly, and more interestingly,
the theory implicitly encoded by a family $P$ of $2$-overlapping fuzzy sets can already be
axiomatized in {\em four-valued {\god} logic}, denoted $\GL_4$: even if $n$ grows ever larger,
it is not necessary to use more than four truth-values. The required background on
finite-valued {\god} logics is recalled in Section \ref{sec:preliminary}.

 This reduction to four-valued {\god} logic continues to hold when we assume that  $P$
 satisfies both the Ruspini condition (\ref{eq:Ruspini}) and the $2$-overlapping condition (\ref{eq:2overlap}).
 Closing a circle of ideas,  in our final Theorem \ref{th:main3} we obtain the axiomatic
 characterization (over $\GL_4$) of those weak Ruspini partitions $P$ that are $2$-overlapping.

\noindent {\bf Acknowledgement.} This paper is a revised and extended version of \cite{ecsqaru07}. We are
grateful to the three anonymous referees for several suggestions that have greatly
improved the presentation of our results. We are also grateful to Stefano Aguzzoli for a useful
conversation on the subject of this paper.

%

\section{Preliminary Results}
\label{sec:preliminary}
 In this section, our aim is twofold. First, we wish to associate with $P$ a
 formula $\alpha_P(X_1, \ldots, X_n)$ in {\god} logic that
encodes all the extra-logical information provided by $P$, as discussed in the introduction.
Second, we wish to explain how
precisely the same information can be encoded in combinatorial terms using appropriate
partially ordered  sets (\emph{posets}, for short). For this, we shall eventually associate
with $P$ (and $\alpha_P$) a poset $F(P)$ (and $F_{{\alpha_P}}$)---see (\ref{eq:F-alpha}) below.

\subsection{{\god}\label{ss:alg} algebras}As a tool, we make use of the algebraic counterpart
of \god{} logic, namely, \emph{\god{} algebras}. These are Heyting
algebras\footnote{For background on Heyting algebras, we refer to \cite{bd}.}
$\langle G, \wedge, \vee, \to, \neg, \top, \bot \rangle$ satisfying the prelinearity
condition $(x\to y) \vee (y \to x) = \top$.
Thus, {\god} algebras are to {\god} logic precisely as Boolean algebras are to classical
propositional logic. The standard correspondence between algebraic and
logical notions generalizes to {\god} logic, and shall be used below.

 The collection of all functions from $[0,1]$ to $[0,1]$ has the structure of a G\"odel algebra under
the following operations,
for $f,g : [0,1] \to [0,1]$.

$\begin{array}{ll}
(f \wedge g) (x) = \min{\{f(x),g(x)\}} &%
(f \vee g)(x) = \max{\{f(x),g(x)\}}\\
(f \to g) (x) = \left\{
\begin{array}{ll}
1 &\hbox{ if }f(x) \leq g(x)\ \ \ \ \ \ \ \\
g(x) &\hbox{ otherwise}
\end{array}
\right.&%
(\neg f)(x)  = \left\{
\begin{array}{ll}
1 &\hbox{ if }f(x) = 0 \\
0 &\hbox{ otherwise.}
\end{array}
\right.%
\end{array}$

The top and bottom elements of the algebra are the constant functions $1$ and $0$, respectively.

We shall denote by $\G{(P)}$  the G\"odel subalgebra of the algebra of all
functions from $[0,1]$ to itself generated by $P$.
 For each integer $k \geq 0$, we
write $\G_k$ for the \emph{free G\"odel algebra} on $k$ free generators $x_1, \ldots, x_k$
corresponding to the propositional variables $X_1, \ldots, X_k$. That is,  $\G_k$   is
the Lindenbaum algebra of the pure {\god} logic restricted to the first $k$ propositional
variables. Then $\G_k$ is finite---it is well-known that {\god} algebras form a locally finite variety of algebras \cite[Theorem 4]{horn}.
Since $\G{(P)}$ is generated by the  $n$ elements $f_1, \ldots, f_n$, there is a congruence $\Theta$ on $\G_n$ such that
the quotient algebra $\G_n / \Theta$ satisfies
\begin{equation}\label{eq:quotient}
\G_n / \Theta \cong \G(P) \ ,
\end{equation}
where $\cong$ denotes isomorphism of G\"odel algebras. We recall that congruences
of a G\"odel  algebra $G$ are in one-one correspondence with
\emph{filters} of $G$, that is, with upward closed subsets closed
under the $\wedge$ operation. In particular, filters of the form
$\uparrow x = \{y \in G\ |\ y \geq x\}$ are called \emph{principal}, as their
corresponding congruences. If, additionally, $G$ is finite, all filters
(and congruences) are necessarily principal.
Therefore, $\Theta$
is generated by a
single equation $\alpha(x_1, \ldots, x_n)$ $=$ $\top$ in the language of  G\"odel algebras. In logical terms, there is a single
formula
\begin{equation}\label{eq:form}
\alpha_P \equiv \alpha_P(X_1, \ldots, X_n)
\end{equation}
over the $n$ propositional variables $X_1, \ldots, X_n$, such that the Lindenbaum algebra of the theory axiomatized
by the single axiom $\alpha_P$ is isomorphic
to $\G(P)$. Note that $\alpha_P$ is uniquely determined by $P$  up to logical
equivalence. Indeed, if $\alpha(X_1,\ldots,X_n)$ is another
formula such that the corresponding equation $\alpha(x_1,\ldots,x_n)=\top$
generates the congruence $\Theta$, then, algebraically,
$\alpha$ and $\alpha_P$ represent the unique element $x$ of $\G_n$ that
generates the unique principal filter $\uparrow x$ corresponding to $\Theta$. Hence,
$\vdash \alpha \leftrightarrow \alpha_P$, where we write $\alpha \leftrightarrow \alpha_P$ as a shorthand for
$(\alpha \rightarrow \alpha_P) \wedge (\alpha_P \rightarrow \alpha)$.

Intuitively, then, the formula $\alpha_P$ encodes \emph{all} relations between the fuzzy
sets $f_1, \ldots, f_n$ that {\god} logic is capable to express. The standard argument
above only grants the existence and uniqueness of $\alpha_P$, given $P$.
We now turn to the problem of describing $\alpha_P$ concretely in terms of $P$.

\subsection{Combinatorial representation}\label{ss:comb}
Any finite Boolean algebra can be thought of as the family of all subsets of a finite set,
endowed with set-theoretic operations. For finite {\god} algebras, one needs to replace sets with \emph{forests}, as follows.

Recall that, given a poset $(F,\leq)$ and a set $Q \subseteq F$, the \emph{downset}
of $Q$ is
$$\downarrow Q = \{x \in F\ |\ x \leq q,\ \mbox{for some}\ q \in Q\}.$$
\noindent We write $\downarrow q$ for $\downarrow \{q\}$. A poset $F$ is a \emph{forest} if for all $q \in F$
the downset $\downarrow q$  is a chain (\textit{i.e.}, a totally ordered set).
A \emph{leaf} is a maximal element of $F$. A \emph{tree} is a forest with a bottom element, called the \emph{root}
of the tree.
A \emph{subforest} of a forest $F$ is the downset of some $Q \subseteq F$. The \emph{height} of a chain is the number of its elements.
The \emph{height} of a forest is the maximum height of any inclusion-maximal chain of the forest.

Let $\Sub{(F)}$ denote the family of all subforests of a forest $F$. It so happens that $\Sub{(F)}$ has a natural structure
of G\"odel algebra, where $\wedge$ and $\vee$ are given by union and intersection of subforests, and implication is defined,
for $F_1,F_2 \in \Sub(F)$, as
\[
F_1 \rightarrow F_2 = \{q \in F\ |\ \downarrow q \cap F_1 \subseteq\ \downarrow q \cap F_2 \}.
\]
The constants $\bot,\top$ are the empty forest and $F$ itself, respectively. Negation is defined
by $\neg F_1 = F_1 \rightarrow \bot$. It turns out that any finite {\god} algebra
is representable as $\Sub{(F)}$, for some choice of $F$ that is unique to within
a poset isomorphism. See \cite[\S 2]{APAL} for a concise treatment and further references.

 The forest $\F_n$ such that $\G_n\cong \Sub{(\F_n)}$ has special importance, as it is
 associated with the pure {\god} logic over the propositional variables  $X_1, \ldots, X_n$.
 We next show how to explicitly describe  $\F_n$ in the elementary language of $[0,1]$-valued
 assignments. This description plays a key role in what follows.

\begin{defn}
\label{def:n-equivalent}
We say that two assignments $\mu$ and $\nu$ are \emph{equivalent over the first
$n$ variables}, or \emph{$n$-equivalent}, written $\mu \equiv_n \nu$,
if and only if there exists a permutation
$\sigma : \underline{n} \to \underline{n}$
such that\textup{:}
\begin{equation}
\label{eq:assignment}
0 \preceq_0 \mu(X_{\sigma(1)}) \preceq_1 \cdots \preceq_{n-1} \mu(X_{\sigma(n)}) \preceq_{n} 1\ ,
\end{equation}
$$0 \preceq_0 \nu(X_{\sigma(1)}) \preceq_1 \cdots \preceq_{n-1} \nu(X_{\sigma(n)}) \preceq_{n} 1\ ,$$
where $\preceq_i\ \in \{<,=\}$, for $i=0,\dots,n$.
\end{defn}
Clearly, $\equiv_n$ is an equivalence relation.
Throughout, we write $\F_n$ for
the (finite) set of equivalence classes of $\equiv_n$.
Here, we are abusing notation in that $\F_n$ already denotes a forest such that
$\G_n\cong \Sub{(\F_n)}$. In fact, (\emph{i}) in Proposition \ref{prop:free} below
shows that our usage is harmless.

 It is not difficult to show that if $\alpha(X_1,\dots,X_n)$ is a formula
in G\"{o}del logic, and $\mu$, $\nu$ are two $n$-equivalent assignments, then
\begin{equation}
\label{eq:assignments}
\mu(\alpha(X_1,\dots,X_n))=1\ \ \mbox{if and only if}\ \ \nu(\alpha(X_1,\dots,X_n))=1.
\end{equation}

 We can further endow  $\F_n$ with a partial order.
\begin{defn}\label{D: order between assignments}
Let $[\mu]_{\equiv_n},[\nu]_{\equiv_n} \in \F_n$,
and let $\sigma : \underline{n} \to \underline{n}$ be a permutation such that
$$0 \preceq_0 \nu(X_{\sigma(1)}) \preceq_1 \cdots \preceq_{n-1} \nu(X_{\sigma(n)}) \preceq_{n} 1\,,$$
$$0\; \widetilde{\preceq}_0\; \mu(X_{\sigma(1)})\; \widetilde{\preceq}_1 \cdots
\widetilde{\preceq}_{n-1}\; \mu(X_{\sigma(n)})\ \widetilde{\preceq}_{n}\; 1\,,$$
where $\preceq_i,\,\widetilde{\preceq}_i\ \in \{<,=\}$, for $i=0,\dots,n$.
We define $[\mu]_{\equiv_n} \leq [\nu]_{\equiv_n}$ if and only if
there exists an index
$k\in\{0,\dots,n\}$
such that
\begin{enumerate}
        \item[\textit{i}\textup{)}] $\widetilde{\preceq}_i$ coincides with $\preceq_i\ $ if $\ 0\leq i \leq k$,
        \item[\textit{ii}\textup{)}] $\widetilde{\preceq}_i$ coincides with $=\ $  if $\ k+1 \leq i \leq n$.
\end{enumerate}
\end{defn}

\begin{exmp}
Let $\mu$, $\nu$, $\xi$ be assignments such that
\begin{itemize}
\item $\mu(X_1)=1$, $\mu(X_2)=1/3$, $\mu(X_3)=0$, $\mu(X_4)=1$,
\item $\nu(X_1)=1$, $\nu(X_2)=1/4$, $\nu(X_3)=0$, $\nu(X_4)=1/2$,
\item $\xi(X_1)=1$, $\xi(X_2)=1/2$, $\xi(X_3)=0$, $\xi(X_4)=1/2$.
\end{itemize}
For $\sigma(1)=3$, $\sigma(2)=2$, $\sigma(3)=4$, $\sigma(4)=1$, one has
\begin{itemize}
\item $0 = \mu(X_3) < \mu(X_2) < \mu(X_4) = \mu(X_1) = 1\,,$
\item $0 = \nu(X_3) < \nu(X_2) < \nu(X_4) < \nu(X_1) = 1\,,$
\item $0 = \xi(X_3) < \xi(X_2) = \xi(X_4) < \xi(X_1) = 1\,.$
\end{itemize}
Thus, according to Definition \ref{D: order between assignments}, $[\mu]_{\equiv_n} \leq [\nu]_{\equiv_n}$,
and $[\xi]_{\equiv_n}$ is uncomparable to both $[\mu]_{\equiv_n}$ and $[\nu]_{\equiv_n}$.
\end{exmp}

One checks that $\leq$ in Definition \ref{D: order between assignments}
indeed is a partial order on $\F_n$, and $(\F_n,\leq)$ is in fact a forest \cite[Lemma 3.3]{ieee}.
Direct
inspection shows that
\begin{enumerate}
\item[a)] the roots of the trees are the equivalence classes of Boolean assignments,
\item[b)] the equivalence class $[\mu]_{\equiv_n}$ such that $\mu(X_1) = \cdots = \mu(X_n) = 0$
is the only tree having height 1, and
\item[c)] the leaves are those equivalence classes of assignments in which no variable is set to $1$.
%
\end{enumerate}

We can now sum up the relationships between finite forests and finite {\god} algebras, as follows.

\begin{figure}[ht!]
        \begin{center}
                \includegraphics{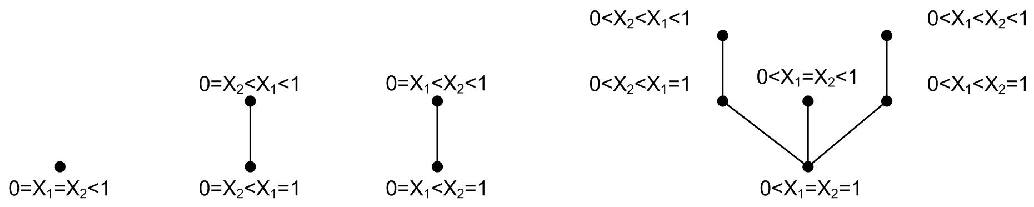}
        \end{center}
        \caption{The forest $\F_2$.}
        \label{F: F2}
\end{figure}

For each $i = 1,\dots,n$, let $\chi_i = \{[\mu]_{\equiv_n}\ |\ \mu(X_i)=1\}$
be the $i^{\textrm{th}}$ \emph{generating subforest} of $\F_n$. We recall that the prime (lattice) filters of a {\god}
algebra $G$ represent precisely those congruences $\Theta$ such that $G/\Theta$ is totally ordered.
\begin{prop}\label{prop:free}  Fix an integer $k \geq 0$.
\textup{(}i\textup{)} $\Sub{(\F_k)}$ is \textup{(}isomorphic to\textup{)} the free  G\"odel algebra on $k$ free generators.
A free generating set is given by the collection of generating subforests.
\textup{(}ii\textup{)} Up to isomorphism, the quotients of $\Sub{(\F_k)}$ are precisely
the algebras of the form $\Sub{(F)}$, for $F \in \Sub{(\F_k)}$. \textup{(}iii\textup{)}
The set of prime filters ordered by reverse inclusion of
$\Sub{(F)}$ is order-isomorphic to $F$ for every $F\in\Sub{(\F_k)}$.
\end{prop}
\begin{pf}
The proof is a straightforward translation of \cite[Remark 2 and Proposition 2.4]{APAL}
in the language of equivalence classes of assignments introduced above.
\end{pf}
Figure \ref{F: F2} shows the forest $\F_2$, whose nodes are labelled by the ordering of
variables under a given assignment as in (\ref{eq:assignment}). However, for the sake of
readability, here and in the following figure we write $X_i$ instead of $\mu(X_i)$.

\subsection{The forest determined by $P$}\label{ss:Pforest}
We can now associate with $P$ a uniquely determined forest.
As an immediate consequence of Proposition \ref{prop:free}, we can reformulate (\ref{eq:quotient}) as follows: $P$ uniquely determines
a congruence $\Theta'$ on $\Sub(\F_n)$, and a subforest $F(P)$ of $\F_n$ such that
\[
\Sub(\F_n) / \Theta' \cong \Sub(F(P)) \cong \G(P)\,.
\]
To relate $\Theta'$ with the formula $\alpha_P$ in (\ref{eq:form}) or, equivalently, with $F(P)$, we shall
give an explicit description of $F(P)$. To this end, it is convenient to introduce the following notion.
\begin{defn}
\label{def:realized at x}
Let $[\mu]_{\equiv_n} \in \F_n$ and $x\in[0,1]$.
We say $[\mu]_{\equiv_n}$ is \emph{realized by $P$ at $x$} if
there exists a permutation $\sigma : \underline{n} \to \underline{n}$ such that
\[
0 \preceq_0 f_{\sigma(1)}(x) \preceq_1 \cdots \preceq_{n-1} f_{\sigma(n)}(x) \preceq_{n} 1\,,
\]
\[
0 \preceq_0 \mu(X_{\sigma(1)}) \preceq_1 \cdots \preceq_{n-1} \mu(X_{\sigma(n)}) \preceq_{n} 1\,,
\]
where $\preceq_i\ \in \{<,=\}$, $i \in \{0,\dots,n\}$.
\end{defn}
\begin{prop}We have
\label{prop:fp}
\begin{equation}\label{eq:prop2}
F(P)=\,\downarrow\{[\mu]_{\equiv_n}\in\F_n\, |\, [\mu]_{\equiv_n}\,\mbox{is realized by}\ P\ \mbox{at some}\,x\in[0,1]\}\ .
\end{equation}
\end{prop}
\begin{pf}
We first construct a subdirect representation of $\G(P)$.
We shall then use Proposition \ref{prop:free} to identify $F(P)$ with the forest
of prime filters of $\G(P)$. This will allow us to prove the desired equality (\ref{eq:prop2}).

To construct the subdirect representation, note that there exists
a finite set $\{x_1,\dots,x_m\} \subseteq[0,1]$
such that for each $y \in [0,1]$, if $[\mu]_{\equiv_n} \in F(P)$ is realized by $P$ at $y$, then it is also realized
by $P$ at $x_i$, for some $i\in\underline{m}$. Moreover, one checks
that evaluating the elements of $\G(P)$ at $x_i$ yields a totally
ordered \god{} algebra $C_{x_i}$ that is a homomorphic image of $\G(P)$ via the quotient map $q_i$ given by restriction to $x_i$.
The homomorphism
$$s: \G(P) \hookrightarrow \prod_{i=1}^{m}C_{x_i}$$
given by
$$g\in\G(P) \longmapsto (q_1(g),\dots,q_m(g))$$
is injective. Indeed, let $g\neq h \in \G(P)$, say $g(y) > h(y)$ for $y\in [0,1]$.
For the sake of brevity, we shall only deal with the case $1 > g(y) > h(y) > 0$. Then
$g(y)=f_i(y)$ and $h(y)=f_j(y)$ for $i\neq j$. Let
$[\mu]_{\equiv_n}$ be the assignment realized by $P$ at $y$.
There exists $u\in\underline{m}$ such that $[\mu]_{\equiv_n}$ is realized by $P$ at $x_u$, and therefore
$f_i(x_u) > f_j(x_u)$, which proves $s(g) \neq s(h)$.

It now follows that $s$ is a subdirect representation of $\G(P)$.
By Proposition \ref{prop:free}(\emph{iii}) we identify prime filters of $\G(P)$ with elements of $F(P)\subseteq\F_n$.
The prime filters that are kernels of $q_1,\dots,q_m$ must comprise all inclusion-minimal prime filters of $\G(P)$,
\emph{i.e.}, all leaves of $F(P)$, for otherwise $s$ could not be a subdirect representation.
Therefore, the classes $[\mu]_{\equiv_n}$ realized by $P$ at some $x\in[0,1]$
comprise all leaves of $F(P)$ (and possibly other elements). Since any forest is the downset of its leaves
the proposition is proved.
\end{pf}
In general, we associate with a formula $\alpha(X_1, \ldots, X_n)$ the uniquely determined subforest of $\F_n$, denoted
$F_\alpha$, as follows:
\begin{equation}\label{eq:falpha}
F_\alpha = \{[\mu]_{\equiv_n} \in \F_n\ |\ \mu(\alpha)=1\}\ .
\end{equation}
By (\ref{eq:assignments}), $F_\alpha$ does not depend on the choice of $\mu$. Clearly, $F_\alpha$ corresponds
to the quotient algebra $\Sub(\F_n) / \Theta'$, where $\Theta'$ is the congruence generated by
$\alpha(X_1,\dots,X_n) = \top$. Finally, by the foregoing we have
\begin{equation}
\label{eq:F-alpha}
F_{\alpha_P} = F(P)\ .
\end{equation}
\subsection{Finite-valued {\god} logics}\label{ss:finitevalued}
In Section \ref{s:mainbis} we are going to deal with four-valued {\god} logic.
Here we provide the needed background. Fix an integer $t\geq 2$, and consider the set
of truth values $T_t=\{0=\frac{0}{t-1}, \frac{1}{t-1}, \ldots, \frac{t-2}{t-1}, \frac{t-1}{t-1}=1\} \subseteq [0,1]$.
To define $n$-valued {\god} logic semantically, we consider the same set of well-formed formulas
over $X_1,X_2, \ldots$ as for $\GL_\infty$, but  we restrict assignments to those taking values
in $T_t$, that is, to \emph{$t$-valued assignments}. A tautology of $t$-valued {\god} logic $\GL_t$
is defined as a formula that takes value $1$ under
any $t$-valued assignment. Syntactically, we need to add one axiom scheme to those of $\GL_\infty$ in order to obtain  a
completeness theorem for $\GL_t$. Namely, consider the axiom
\begin{equation*}
\alpha_1\vee(\alpha_1\to\alpha_2)\vee\cdots\vee(\alpha_1\wedge\cdots\wedge\alpha_{t-1} \to \alpha_{t})\,.
\tag{$\textsc{Lin}_t$}
\end{equation*}
Using \textit{modus ponens} as the only deduction rule, one proves that the
axioms of $\GL_\infty$ together with $(\textsc{Lin}_t)$ provide a complete
axiomatization\footnote{Readers interested in proof-theoretic aspects of
{\god} logics are referred to \cite{bcf} for an extensive discussion with further references.} of $\GL_t$. We write
$
 \vdash_{\GL_t}\alpha
$
to mean that the formula $\alpha$ is provable in $\GL_t$.

It is straightforward to extend to $\GL_t$ the combinatorial representation theory of
Subsection \ref{ss:comb}. For this, we use partially ordered equivalence classes of
assignments as in Definitions \ref{def:n-equivalent} and \ref{D: order between assignments},
except that we only consider  $t$-valued assignments. Contemplation of the
meaning of ($\textsc{Lin}_t$) shows that
the forest $\F_n^t$ associated with the pure $t$-valued {\god} logic $\GL_t$ is
order-isomorphic to the subforest of $\F_n$ consisting of all elements having
height at most $t-1$. In other words, truncating $\F_n$ to height $t-1$ yields
$\F_n^t$.  The correspondence for $\GL_\infty$ between subforests, formulas, and
quotient algebras given by the foregoing now extends to $\GL_t$ in the obvious manner.

\section{\god{} Approximation of Ruspini Partitions}\label{s:main}
%
Let $P$ be a Ruspini partition. It is clear that those
assignments $\mu$ to $X_1,\dots,X_n$ such that either $\mu(X_i)=0$, for all $i \in \underline{n}$, or $\mu(X_i)<1$
for exactly one index $i$, and $\mu(X_j)=0$, for all $j \neq i$, cannot evaluate $\alpha_P$ to $1$.
Equivalently, these assignments cannot be realized
by $P$ at any $x \in [0,1]$. The following definition isolates a class of subforests $\Rus_n \subseteq \F_n$
that omits from $\F_n$ precisely those points corresponding to such assignments.

\begin{defn}\label{def:ruspini}We denote by $\Rus_n$ the subforest of $\F_n$ obtained by removing from $\F_n$
the single tree having height $1$, and the leaves of all the trees having height $2$.
We call $\Rus_n$ the \emph{Ruspini forest}.
\end{defn}
\begin{figure}[ht!]
        \begin{center}
                \includegraphics{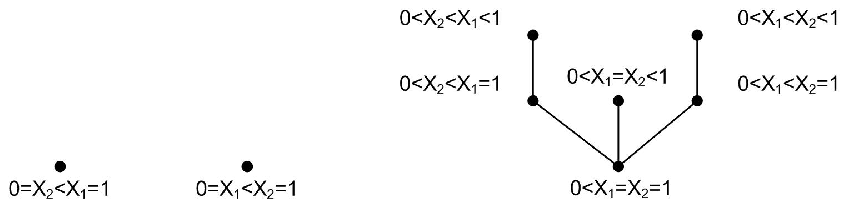}
        \end{center}
        \caption{The Ruspini forest $\Rus_2$.}
        \label{F: R2}
\end{figure}
We now show how to explicitly axiomatize $\Rus_n$.
\begin{defn}We define the \emph{Ruspini axiom} $\rho_n = \alpha \vee \beta$,
where
$$\alpha = \bigvee_{1 \leq i < j \leq n}(\neg\neg X_i \wedge \neg\neg X_j)\,,\ \ \mbox{and}
\ \ \ \beta = \bigvee_{1 \leq i \leq n} (X_i \wedge \bigwedge_{1 \leq j \neq i \leq n} \neg X_j)\,.$$
\end{defn}
Recall that the formula $\rho_n$ uniquely determines a subforest $F_{\rho_n}\subseteq \F_n$ as in (\ref{eq:falpha}). In fact:
\begin{prop}
$F_{\rho_n}  = \Rus_n \,.$
\end{prop}
\begin{pf}
Fix an assignment $\mu$. Since
$$\mu(\neg\neg X)  = \left\{
\begin{array}{ll}
0 &\hbox{ if }\mu(X) = 0 \\
1 &\hbox{ otherwise,}
\end{array}
\right.$$
$\mu(\alpha)\neq 1$ if and only if at most one variable $X_{i_0}$ satisfies
$\mu(X_{i_0})\neq 0$.

Observe now that $\mu(\beta) = 1$ if and only if there exists $i \in \underline{n}$ such that, for
$j\neq i,\ \mu(X_{i})=1$ and $\mu(X_j)=0$.

Therefore,
$\mu(\rho_n)=\mu(\alpha \vee \beta) \neq 1$ if and only if there exists $i_0 \in \underline{n}$ such that, for
$j\neq i_0,\ \mu(X_{i_0})<1$ and $\mu(X_j)=0$.
It is now straightforward to verify that the latter condition
holds if and only if $[\mu]_{\equiv_n} \notin \rus_n$.
\end{pf}

Let us introduce a property of $P$ that we shall use in our main result.
Let $\lambda: [0,1] \to [0,1]$ be an order preserving map such that $\lambda(0)=0$ and $\lambda(1)=1$, and let
$t= \inf{\lambda^{-1}(1)}$. If the restriction of $\lambda$ to $[0,t]$ is an order isomorphism
between $[0,t]$ and $[0,1]$, we say $\lambda$
is a \emph{comparison map}.

\begin{defn}\label{def:weak ruspini}We say $P$ is a  \emph{weak Ruspini partition} if for all $x \in [0,1]$,
there exist $y \in [0,1]$, a comparison map $\lambda$, and an order isomorphism $\gamma$ from $[0,1]$ to itself, such
that
\begin{enumerate}
        \item[\textup{(}i\textup{)}] $\lambda(f_i(y)) = f_i(x)$, for all $i \in \underline{n}$.
        \item[\textup{(}ii\textup{)}] $\sum_{i=1}^{n}{\gamma(f_i(y))}=1$.
\end{enumerate}
\end{defn}

\begin{exmp} The set of functions $P=\{f_1,f_2\}$ shown in Figure \textup{\ref{fig:weakruspini}}
is a weak Ruspini partition.
\begin{figure}[ht!]
        \begin{center}
                \includegraphics{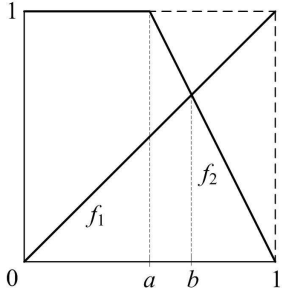}
        \end{center}
        \caption{A weak Ruspini partition $\{f_1,f_2\}$.}
\label{fig:weakruspini}
\end{figure}
Indeed, for $x=0$ or $x=1$, conditions \textup{(}\textit{i}\textup{)} and
\textup{(}\textit{ii}\textup{)} in Definition \ref{def:weak ruspini}
are trivially satisfied with $y=x$, and $\lambda$ and $\gamma$ the identity functions.
\textup{(}More generally, for all $x \in [0,1]$ where the Ruspini condition locally holds as $\sum_{i=1}^{n}{f_i(x)}=1$,
\textup{(}\textit{i}\textup{)} and \textup{(}\textit{ii}\textup{)} are satisfied in this manner.\textup{)}
If $x \in (a,1)$ we can still choose $y=x$ and $\lambda$
the identity function. Then, since the values of $f_1$, $f_2$ at $x$ satisfy $0<f_1(x),f_2(x)<1$,
it is clear that there is an order isomorphism $\gamma$ that shifts this values to $0<\gamma(f_1(x)),\gamma(f_2(x))<1$
so that $\gamma(f_1(x))+\gamma(f_2(x))=1$. Finally, consider $x \in (0,a]$. Here, $x=y$ does not work, because
then $0<f_1(x)<f_2(x)=1$, and regardless of our choice of $\gamma$ we have
$0<\gamma(f_1(x))<\gamma(f_2(x))=1$, whence $\gamma(f_1(x))+\gamma(f_2(x))>1$.
However, let $y \in (a,b)$. Then, we can choose $\lambda$ such that
$\lambda(f_1(y))=f_1(x)$, and the restriction of $\lambda$ to  $[0,f_2(y)]$ is an
order isomorphism onto $[0,1]$---hence $\lambda(f_2(y))=f_2(x)$, too. In particular $\lambda$ carries  $[f_2(y),1]$
to $1$. For this choice of $\lambda$, \textup{(}\textit{i}\textup{)} is satisfied. As before, it is
easy to construct an order isomorphism $\gamma$ satisfying \textup{(}\textit{ii}\textup{)} with respect to
our chosen $y$. \textup{(}Thus, $\lambda$ preserves
the relative order of the values of $f_i$, except that it can collapse the values above
a fixed $t \in (0,1]$ to $1$. Then the Ruspini condition is to be satisfied by the values at
$y$.\textup{)}
\end{exmp}

The importance of comparison maps to our purposes is brought out by our next result.
The following lemma relates the order between points of $\F_n$ realized by any $P$
with the existence of an appropriate comparison map. Further, it relates the existence
of leaves of $\Rus_n$ realized by $P$ with the existence of an appropriate order isomorphism
of the real unit interval.

\begin{lem}
\label{l:comparisonmap}
Let $[\mu]_{\equiv_n},\ [\nu]_{\equiv_n} \in \F_n$ and $x,y \in [0,1]$ such that
$[\mu]_{\equiv_n}$ and $[\nu]_{\equiv_n}$ are realized by $P$ at $x$ and $y$, respectively.
Then the following are equivalent.\vspace{-1mm}
\begin{enumerate}
        \item[\textup{(}i\textup{)}] $[\mu]_{\equiv_n} \leq [\nu]_{\equiv_n}$.
        \item[\textup{(}ii\textup{)}] There exists a comparison map $\lambda:[0,1] \to [0,1]$
                        with $\lambda(f_i(y)) = f_i(x)$, for all $i \in \underline{n}$.
\end{enumerate}
\vspace{-1mm}Moreover, the following are equivalent.\vspace{-1mm}
\begin{enumerate}
        \item[\textup{(}iii\textup{)}] $[\mu]_{\equiv_n}$ is a leaf of $\Rus_n$.
        \item[\textup{(}iv\textup{)}] There exists an order isomorphism $\gamma:[0,1] \to [0,1]$
                        with $\sum_{i=1}^{n}{\gamma(f_i(x))}=1$.
\end{enumerate}
\end{lem}
\begin{pf}
(\emph{i})$ \Rightarrow $(\emph{ii}).
By Definitions \ref{D: order between assignments} and \ref{def:realized at x},
there exists a permutation $\sigma: \underline{n} \to \underline{n}$
such that\\
\centerline{$0 \preceq_0 f_{\sigma(1)}(y) \preceq_1 \cdots \preceq_{n-1} f_{\sigma(n)}(y) \preceq_{n} 1\,,$}
\centerline{$0\; \widetilde{\preceq}_0\; f_{\sigma(1)}(x)\; \widetilde{\preceq}_1 \cdots \widetilde{\preceq}_{n-1}\; f_{\sigma(n)}(x)\ \widetilde{\preceq}_{n}\; 1\,,$}
where $\preceq_i,\,\widetilde{\preceq}_i \in \{<,=\}$, and there is $k\in\{0,\dots,n\}$
satisfying \emph{i}) and \emph{ii}) in Definition \ref{D: order between assignments}.
We deal with the case $k < n$ only; the case $k=n$ is a trivial variation thereof.
We define
$\Lambda$ by $\Lambda(f_{\sigma(i)}(y))=f_{\sigma(i)}(x)$, for $1\leq i \leq k$, and
$\Lambda(f_{\sigma(i)}(y))= 1$ if $k+1 \leq i \leq n$. We extend $\Lambda$ to a comparison map
as follows.
Consider the closed intervals $I_0=[0,f_{\sigma(1)}(y)]$, $J_0=[0,f_{\sigma(1)}(x)]$,
$I_i=[f_{\sigma(i)}(y),f_{\sigma(i+1)}(y)]$ and $J_i=[f_{\sigma(i)}(x),f_{\sigma(i+1)}(x)]$,
for $1 \leq i \leq k$. Now let us fix $0 \leq h \leq k$. Note that if $I_h$ collapses to a point,
then $J_h$ also collapses to a point. Therefore,
in all cases we can choose
order isomorphisms $\lambda_h: I_h \to J_h$. Moreover, set $I_{k+1}=[f_{\sigma(k+1)}(y), 1]$ and $\lambda_{k+1}: I_{k+1} \to \{1\}$.
Since $\lambda_h$ and $\lambda_{h+1}$ agree at $I_h \cap I_{h+1}$ by construction,
the function $\lambda: [0,1] \to [0,1]$ defined by $\lambda(r) = \lambda_j(r)$ if $r\in I_j$, for $0 \leq j \leq k+1$, is
a comparison map satisfying (\emph{ii}).

(\emph{ii})$ \Rightarrow $(\emph{i}).  Immediate from Definitions \ref{D: order between assignments} and \ref{def:realized at x}.

(\emph{iii})$ \Rightarrow $(\emph{iv}). It is an exercise to check that $[\mu]_{\equiv_n}$ is a leaf of $\Rus_n$ if and
only if exactly one of the following two cases hold.

\noindent\emph{Case 1}. There exists $i_0$ such that $\mu(X_{i_0})=1$ and $\mu(X_i)=0$ for $i\neq i_0$.\\
Let $\gamma$ be the identity map. By Definition \ref{def:realized at x}, we have
$\sum_{i=1}^{n}{\gamma(f_i(x))}=1$.

\noindent\emph{Case 2}. For all $i$, $\mu(X_i)<1$, and there exist $i_0,i_1$ such that $0<\mu(X_{i_0})\leq\mu(X_{i_1})$.\\
Let us write\\
\centerline{$0 \preceq_0 f_{\sigma(1)}(x) \preceq_1 \cdots \preceq_{n-1} f_{\sigma(n)}(x) \preceq_{n} 1\,,$}
for some permutation $\sigma$ and $\preceq_i\in\{<,=\}$.
We shall assume $\preceq_0$ is $<$. The case where some $f_i$ takes value zero at $x$ is entirely similar.

Now consider the  $(n-1)$-dimensional simplex\footnote{For all
unexplained notions in combinatorial topology, please see \cite{rourke}.} $S_n$,
given by the convex hull of the standard basis of $\R^n$. Let $S^{(1)}_n$ be the
simplicial complex given by the first barycentric subdivision of $S_n$. The $(n-1)$-dimensional simplices
of $S^{(1)}_n$ are in bijection with the permutations of $\underline{n}$, and  the solution set of
the inequalities
\begin{equation}
\label{eq:S}
0 \leq r_1 \leq \cdots \leq r_n \leq 1
\end{equation}
in $S^n$ is an $(n-1)$-dimensional simplex $S\in S^{(1)}_n$.
Consider the equalities
\begin{equation}
\label{eq:eq}
r_i = r_{i+1}
\end{equation}
for each $i=1,\dots,n-1$  such that $\preceq_i$ is $=$. Then the solution set of
(\ref{eq:S}) and (\ref{eq:eq}) is a nonempty face $T$ of
$S$.
Consider next the strict inequalities
\begin{equation}
\label{eq:ineq}
\left\{\begin{array}{ll}
r_i < r_{i+1} \\
0<r_1\\
r_n<1
\end{array}
\right.
\end{equation}
for all $i=1,\ldots,n-1$ such that $\preceq_i$ is $<$. Then the solution set of
(\ref{eq:S}), (\ref{eq:eq}), and (\ref{eq:ineq}) is the relative interior $T^\circ$ of $T$. Since $T$ is nonempty,
$T^\circ$ is nonempty. The barycenter $b=(b_1,\dots,b_n)$ of $T$ lies in $T^\circ$. Since $b\in S_n$,
we have $\sum_{k=1}^{n}b_k=1$. Moreover, by construction,\\
\centerline{$0 \preceq_0 b_1 \preceq_1 \cdots \preceq_{n-1} b_n \preceq_{n} 1\,.$}
We define $\Gamma$ by $\Gamma(f_{\sigma(i)})=b_i$. Arguing as in the proof of (\emph{i})$\Rightarrow$(\emph{ii}),
we conclude that there is an extension of $\Gamma$ to an order isomorphism $\gamma:[0,1]\to[0,1]$
satisfying (\emph{iv}).

(\emph{iv})$ \Rightarrow $(\emph{iii}). Suppose $[\mu]_{\equiv_n}$ is not a leaf of $\Rus_n$.
Thus, exactly one of the following two cases holds.

\noindent\emph{Case 1}. $[\mu]_{\equiv_n}\in\F_n\setminus\Rus_n$.\\
In this case there exists $i_0$ such that $\mu(X_{i_0})<1$
and $\mu(X_i)=0$  for $i\neq i_0$. Using Definition \ref{def:realized at x}, we have
$\sum_{i=1}^{n}{\gamma(f_i(x))}<1$, for each order isomorphism $\gamma$.

\noindent\emph{Case 2}. $[\mu]_{\equiv_n}\in\Rus_n$, but $[\mu]_{\equiv_n}\in\Rus_n$ is not a leaf of $\Rus_n$.\\
It is easy to check that there exist $i_0,i_1$ such that $0<\mu(X_{i_0})\leq\mu(X_{i_1})=1$. Using Definition \ref{def:realized at x},
we have $f_{i_1}(x)=1$ and $f_{i_0}(x)>0$, and thus $\sum_{i=1}^{n}{\gamma(f_i(x))}>1$, for each order isomorphism $\gamma$.
\end{pf}

To state our main result we still need to show how to obtain a formula $\psi_{[\mu]_{\equiv_n}}$
associated with a given element $[\mu]_{\equiv_n} \in \F_n$ such that
$\psi_{[\mu]_{\equiv_n}}$ evaluates to $1$ exactly on $\downarrow[\mu]_{\equiv_n}$.
To this end, we define the derived connective $\alpha \triangleleft \beta = ((\beta \to \alpha)\to \beta)$.
Given an assignment $\mu$ we have that
$$\mu(\alpha \triangleleft \beta)  = \left\{
\begin{array}{ll}
1 &\hbox{ if }\mu(\alpha) < \mu(\beta)\ \hbox{or}\ \mu(\alpha)=\mu(\beta)=1\\
\mu(\beta) &\hbox{ otherwise.}
\end{array}
\right.$$
Suppose now that, for a given permutation
$\sigma : \underline{n} \to \underline{n}$,
$$0 \preceq_0 \mu(X_{\sigma(1)}) \preceq_1 \cdots \preceq_{n-1} \mu(X_{\sigma(n)}) \preceq_{n} 1\,,$$
where $\preceq_i\,\in \{<,=\}$, $i=0,\dots,n$. We associate to $[\mu]_{\equiv_n}$ the formula
\begin{equation}\label{eq:normalform}\psi_{[\mu]_{\equiv_n}}=(\bot\ \bowtie_0\ X_{\sigma(1)})\wedge(X_{\sigma(1)}\ \bowtie_1 \
X_{\sigma(2)})\wedge\cdots\wedge(X_{\sigma(n)}\ \bowtie_{n}\ \top)\,,\end{equation}
where $\bowtie_i = \triangleleft$ if $\preceq_i$ is $<$, and $\bowtie_i = \leftrightarrow$ otherwise.
\begin{lem}
\label{l:normalform}
$F_{\psi_{[\mu]_{\equiv_n}}}=\ \downarrow[\mu]_{\equiv_n}$.
\end{lem}
\begin{pf}
We omit the straightforward verification. Compare \cite{baaz}, where a
theory of \emph{chain normal forms} for {\god} logic is introduced using similar tools.
\end{pf}

Given a forest $F \subseteq \F_n$, let us indicate with $\Leaf(F)$ the set of leaves of
$F$.

\begin{lem}
\label{l:normalformforest}
Fix a forest $F \subseteq \F_n$, and let $\alpha(X_1,\ldots,X_n)$ be
a formula as in \textup{(\ref{eq:falpha})} such that $F_\alpha = F$.
Then
\begin{equation}\label{eq:F-alphanormalform}
\vdash \alpha \leftrightarrow \bigvee_{l\in\Leaf(F_\alpha)}\psi_l\,.
\end{equation}
\end{lem}
\begin{pf}
Set $\beta = \bigvee_{l\in\Leaf(F_\alpha)}\psi_l$. Then, by the definition of $\leftrightarrow$,
(\ref{eq:F-alphanormalform}) holds if and only if $F_\alpha = F_\beta$.
But, by Lemma \ref{l:normalform} along with the definition of $\vee$, $F_\beta$
is the downset of the leaves of $F_\alpha$, whence it coincides with $F_\alpha$.
\end{pf}

Note that, in particular, Lemma \ref{l:normalformforest} yields the promised
explicit construction of $\alpha_P$, for any family of fuzzy sets $P$. Indeed,
using (\ref{eq:F-alpha}),
\begin{equation}\label{eq:F-alphaPnormalform}
\vdash \alpha_P \leftrightarrow \bigvee_{l\in\Leaf(F(P))}\psi_l\,.
\end{equation}

\begin{defn}\label{d:rsubforest}
We say that a forest $F$ is a \emph{Ruspini subforest} if $F\subseteq\Rus_n$ and
each leaf of $F$ is a leaf of $\Rus_n$.
\end{defn}

\noindent
We can finally prove our first main result.\footnote{In
\cite[p.\ 170]{ecsqaru07}
a different version of this theorem appears, where the formula in (\textit{iii}) regrettably contains a mistake.
}

\begin{thm}
\label{th:main}For any choice of $P$ the following are equivalent.
\begin{enumerate}
\item[\textup{(}i\textup{)}] $P$ is a weak Ruspini partition.
\item[\textup{(}ii\textup{)}] $F(P)$ is a Ruspini subforest.
\item[\textup{(}iii\textup{)}] $\GL_\infty$ proves
\begin{equation}\label{eq:formulainfinito}
\alpha_P \leftrightarrow \bigvee_{{l\in\Leaf(F_{\alpha_P})\cap\Leaf(\Rus_n)}}\psi_l\,.
\end{equation}
\end{enumerate}
Moreover, for any Ruspini subforest $F$ there exists a
Ruspini partition $P'=\{f'_1,\dots,f'_n\}$, with $f'_i:[0,1]\to[0,1]$, such that $F(P') = F$.
\end{thm}
\begin{pf}
Recall from (\ref{eq:F-alpha}) that $F_{\alpha_P}=F(P)$. We tacitly use
the latter identification in the proof below.

(\emph{i}) $\Rightarrow$ (\emph{ii}). By Lemma \ref{l:comparisonmap},
we can reformulate Definition \ref{def:weak ruspini} in terms of assignments as follows.
For all $[\mu]_{\equiv_n}\in\F_n$ realized by $P$ at some $x\in[0,1]$, there exists
$[\nu]_{\equiv_n}\geq[\mu]_{\equiv_n}$ realized by $P$ at some $y\in[0,1]$ such that $[\nu]_{\equiv_n}$
is a leaf of $\Rus_n$. Thus, by Proposition \ref{prop:fp}, $F(P)$ is exactly the downset
of those leaves of $\Rus_n$ realized by $P$ at some $x\in[0,1]$.

(\emph{ii}) $\Rightarrow$ (\emph{iii}).
By Definition \ref{d:rsubforest}, each leaf $k \in \Leaf(F_{\alpha_P})$ is a leaf of $\Rus_n$.
Hence, $\Leaf(F_{\alpha_P})\cap\Leaf(\Rus_n) = \Leaf(F_{\alpha_P})$, and the result follows
from (\ref{eq:F-alphaPnormalform}).

 (\emph{iii}) $\Rightarrow$ (\emph{i}).
Suppose $P$ is not a  weak Ruspini partition.
By Definition \ref{def:weak ruspini}, using Lemma \ref{l:comparisonmap},
there exists $k \in F_{\alpha_P}$ such that one of the following two conditions holds.
\begin{enumerate}
\item[(\textit{a})] $k\in\F_n\setminus\Rus_n$.
\item[(\textit{b})] $k\in\Rus_n$ is a maximal element of $F_{\alpha_P}$, but it is not a leaf of $\Rus_n$.
\end{enumerate}

\noindent We will show that both (\textit{a}) and (\textit{b})
lead to a contradiction. To this purpose, set $\beta = \bigvee_{l\in\Leaf(F_{\alpha_P})\cap\Leaf(\Rus_n)}\psi_l$.
Then $F_\beta$, the forest associated with $\beta$ via (\ref{eq:falpha}), is a subforest of $\Rus_n$. Indeed, by Lemma
\ref{l:normalform}, $F_\beta$ is the downset of those $l \in \F_n$ satisfying $l\in\Leaf(F_{\alpha_P})\cap\Leaf(\Rus_n)$.
By the definition of $\leftrightarrow$,
(\ref{eq:formulainfinito}) holds if and only if $F_{\alpha_P} = F_\beta$. Suppose (\emph{a}) holds.
Then $k$ is an element of $F_{\alpha_P}$ lying strictly above a leaf of $\Rus_n$.
Since, as just shown,  $F_\beta \subseteq \Rus_n$, we infer $F_{\alpha_P} \neq F_\beta$,  a contradiction.
Next suppose  (\textit{b})
holds.
We claim $k \notin F_\beta$. Indeed, since $F_\beta$ is the downset of those $l \in \F_n$
satisfying $l\in\Leaf(F_{\alpha_P})\cap\Leaf(\Rus_n)$,
$k \in F_\beta$ if and only if there exists such an $l$ satisfying $l \geq k$. Since $k \notin \Leaf(\Rus_n)$ by (\textit{b}),
we have $l > k$. Since $l \in \Leaf(F_{\alpha_P})$, the latter inequality means that $k$ is not a leaf of $F_{\alpha_P}$,
a contradiction. We conclude $k \notin F_\beta$, whence  $F_{\alpha_P} \neq F_\beta$, as was to be shown.

 Finally, we prove the last statement of the theorem.
Let $[\mu_1]_{\equiv_n},\dots,[\mu_m]_{\equiv_n}$ be the leaves of $F$.
Partition the interval $[0,1]$ into $m$ intervals $I_1=[0,x_1]$, $I_2=(x_1,x_2]$,$\dots$, $I_m=(x_{m-1},1=x_m]$.
We construct the functions $f'_i$ as follows.
For $i\in\underline{n}$, $j\in\underline{m}$, we set $f'_i(x) = C_{ij} \in \R$ if $x\in I_j$. The constants
$C_{ij}$ are chosen so that
\begin{enumerate}
\item[(\textit{c})] $[\mu_j]_{\equiv_n}$ is realized by $P'$ at $x_j$,
\item[(\textit{d})] $\sum_{i=1}^n C_{ij}=1$.
\end{enumerate}
Obviously, it is always possible to choose $C_{ij}$ so that (\emph{c}) holds.
The proof of (\emph{iii}) $\Rightarrow$ (\emph{iv}) in Lemma \ref{l:comparisonmap}
shows that, in fact, it is always possible to choose $C_{ij}$
so that both  (\emph{c}) and (\emph{d}) hold.
\end{pf}
As a first corollary, we can count the number of Ruspini partitions with $n$ fuzzy sets that can be told apart
by {\god} logic.
In \cite[Theorem 3]{aguzzoli} it is shown that the number of leaves of $\F_n$ is
\begin{equation}\label{eq:Ln}
L_n = 2\sum_{k=1}^n{k!\genfrac{\{}{\}}{0pt}{}{n}{k}}\,,
\end{equation}
where $\genfrac{\{}{\}}{0pt}{}{n}{k}$ is the number of partitions of an
$n$-element set into $k$ classes, \emph{i.e.}, the \emph{Stirling number of the second kind}.
The number $\sum_{k=1}^n{k!\genfrac{\{}{\}}{0pt}{}{n}{k}}$
is the $n^{\textrm{th}}$ \emph{ordered Bell number}, \emph{i.e.}, the number of all ordered partitions of $\underline{n}$.
Compare sequence $A000670$ in \cite{sloane}.

Consider $P' = \{f'_1,\dots,f'_n\}$, where $f'_i:[0,1]\to[0,1]$.
In the light of Section \ref{sec:preliminary}, let us say that $P'$ is \emph{\god -equivalent} to $P$ if $F(P)=F(P')$,
or, equivalently,
$\vdash \alpha_P \leftrightarrow \alpha_{P'}$. Then:

\begin{cor}
\label{cor:enum}
The number of equivalence classes of \god-equivalent weak Ruspini partitions of $n$ elements is $2^{L_n-1}-1$,
where $L_n$ is given by \textup{(\ref{eq:Ln})}.
\end{cor}
\begin{pf}
A weak Ruspini partition $P$ is characterized, up to \god-equi\-va\-lence, by the forest $F(P)$, and therefore
by a subset of leaves of $\Rus_n$. Noting that the number of leaves of $\Rus_n$ is $L_n-1$,
and that for every weak Ruspini partition $P$, $F(P)\neq\emptyset$, the corollary follows.
\end{pf}

Our second corollary deals with continuity. Since implication in {\god} logic has  a
discontinuous semantics, it is impossible to force continuity of all functions of a Ruspini partition
(up to {\god}-equivalence). However, it is always possible to bound the number of discontinuities:

\begin{cor}
\label{cor:stepfunction}
\textup{(}i\textup{)}
There is a Ruspini subforest $F$ such that whenever $F(P)=F$ then each $f_i\in P$
has a point of discontinuity.
\textup{(}ii\textup{)} For all Ruspini subforests $F$ with $L$ leaves there is
a choice of a Ruspini partition $P'=\{f'_1,\dots,f'_n\}$, with
$F(P')=F$ such that each $f'_i:[0,1]\to[0,1]$ has at most
$L-1$ points of discontinuity.
\end{cor}
\begin{pf}
(\emph{i}) It suffices to choose $F\subseteq\Rus_n$ as the forest of all Boolean assignments
which are leaves of $\Rus_n$. (\emph{ii}) The construction used in the proof of the last
statement of Theorem \ref{th:main} yields the desired $P'$.
\end{pf}

%
\section{Four-valued {\god} Logic, and $2$-overlapping Ruspini Partitions}\label{s:mainbis}
%
Following the same outline of the previous section, we now investigate
 how {\god} logic expresses the $2$-overlapping property
of the family $P$ of fuzzy sets.

\begin{defn}\label{def:non-overlapping forest}We denote by $\Nover_n$ the subforest
of $\F_n$ obtained by removing from $\F_n$
all the trees of height $>3$.
\end{defn}
\begin{rem} $\Nover_n$ is the subforest of all equivalence classes of assignments $[\mu]_{\equiv_n}\in\F_n$
such that for all $i\neq j \neq k \in \underline{n}$, at least one of
$\mu(X_{i}) = 0$, $\mu(X_j) = 0$, $\mu(X_k) = 0$, holds.
\end{rem}

We can immediately show how to axiomatize $\Nover_n$.

\begin{defn}\label{def:non-overlapping axiom}
We define the \emph{$2$-overlapping axiom} $\tau_n$ by
$$\tau_n = \bigwedge_{1 \leq i < j < k \leq n}\neg (X_i \wedge X_j \wedge X_k)\,.$$
\end{defn}

\begin{lem}\label{l:Tn}
$F_{\tau_n}  = \Nover_n \,.$
\end{lem}
\begin{pf}
Fix an assignment $\mu$.
Note that $\mu(\tau_n)\neq 1$ if and only if there exist $i\neq j \neq k \in \underline{n}$ such that
$\mu(X_{i})>0$, $\mu(X_j) > 0$, and $\mu(X_k) > 0$.
It is now straightforward to verify that the latter condition
holds if and only if $[\mu]_{\equiv_n} \notin \Nover_n$.
\end{pf}
\begin{lem}
\label{l:ginf to g4}
For any choice of $P$,
$$\vdash_{\GL_4}\,\alpha_P\rightarrow\tau_n\ \ \ \mbox{if and only if}\ \ \ \vdash\,\alpha_P\rightarrow\tau_n\,.$$
\end{lem}
\begin{pf}
($\Leftarrow$) Trivial.

($\Rightarrow$) The formula $\alpha_P\rightarrow\tau_n$ is a tautology of $\GL_4$ if and only if
\begin{equation}\label{eq:lemma g4}
F_{\alpha_P} \cap \F_n^4 \subseteq F_{\tau_n} \cap \F_n^4\,.
\end{equation}
Since, by Lemma \ref{l:Tn}, $F_{\tau_n}=\Nover_n$,
and since $\Nover_n \subseteq \F_n^4$ by direct inspection,
Condition (\ref{eq:lemma g4}) is equivalent to
\begin{equation}\label{eq:lemma g4 B}
F_{\alpha_P} \cap \F_n^4 \subseteq \Nover_n\,.
\end{equation}
We show $F_{\alpha_P} \subseteq \Nover_n$. Suppose there exists $[\mu]_{\equiv_n}\in F_{\alpha_P}$
such that $[\mu]_{\equiv_n}\notin \Nover_n$ ({\it absurdum hypothesis}). By
(\ref{eq:lemma g4 B}) we have $[\mu]_{\equiv_n}\in F_{\alpha_P} \setminus \F_n^4$. Therefore,
the class $[\mu]_{\equiv_n}$ must belong to a tree of $\F_n$ of height $> 3$.
If $[\nu]_{\equiv_n}$ is the root of such tree,
we have $[\nu]_{\equiv_n} \in F_{\alpha_P} \cap \F_n^4$ but,
by Definition \ref{def:non-overlapping forest},  $[\nu]_{\equiv_n}\notin \Nover_n$.
This contradicts (\ref{eq:lemma g4 B}), and our claim is settled.
Thus, $F_{\alpha_P} \subseteq \Nover_n$
and the formula $\alpha_P\rightarrow\tau_n$ is a tautology of $\GL_\infty$.
\end{pf}

Using the preceding lemma, we can now prove:
\begin{thm}
\label{th:main2}For any choice of $P$, the following are equivalent.
\begin{enumerate}
\item[\textup{(}i\textup{)}] $P$ is a $2$-overlapping family.
\item[\textup{(}ii\textup{)}] $F(P)$ is a subforest of $\Nover_n$.
\item[\textup{(}iii\textup{)}] $\vdash_{\GL_4}\,\alpha_P\rightarrow\tau_n$.
\end{enumerate}
\end{thm}
\begin{pf}
(\emph{i}) $\Rightarrow$ (\emph{ii}). All $[\mu]_{\equiv_n}\in\F_n$ realized by $P$ at some
$x\in[0,1]$ are such that for all $i\neq j \neq k \in \underline{n}$, at least one of
$\mu(X_{i}) = 0$, $\mu(X_j) = 0$, $\mu(X_k) = 0$, holds. Thus, $F(P)$ is subforest of $\Nover_n$.

 (\emph{ii}) $\Rightarrow$ (\emph{iii}). Since $F(P) \subseteq \Nover_n$,
we have $\vdash\,\alpha_P\rightarrow\tau_n$.
But then $\vdash_{\GL_4}\,\alpha_P\rightarrow\tau_n$.

 (\emph{iii}) $\Rightarrow$ (\emph{i}). Suppose $P$ is not a $2$-overlapping
family ({\it absurdum hypothesis}).
In other words, suppose that there exist $i\neq j \neq k \in \underline{n}$,
and $x \in [0,1]$, such that
$f_i(x) >0$, $f_j(x) > 0$, and $f_k(x) > 0$.
Thus, there exists $[\mu]_{\equiv_n} \in \F_n$ realized by $P$ at $x$, such that
$\mu(X_{i})>0$, $\mu(X_j) > 0$, and $\mu(X_k) > 0$.
Using (\ref{eq:F-alpha}), $[\mu]_{\equiv_n} \in F_{\alpha_P}$. Clearly,
$[\mu]_{\equiv_n} \notin F_{\tau_n}$. Therefore, $\alpha_P\rightarrow\tau_n$
is not a tautology of $\GL_\infty$. By Lemma \ref{l:ginf to g4}, $\alpha_P\rightarrow\tau_n$
is not a tautology of $\GL_4$. This contradicts (\emph{iii}) and completes the proof.
\end{pf}

Our final aim is to combine Theorems \ref{th:main} and \ref{th:main2}, that is,
to axiomatize $2$-overlapping weak Ruspini partitions in four-valued  {\god} logic.

\begin{thm}
\label{th:main3}For any choice of $P$ the following are equivalent.
\begin{enumerate}
\item[\textup{(}i\textup{)}] $P$ is a $2$-overlapping weak Ruspini partition.
\item[\textup{(}ii\textup{)}] $F(P)$ is a Ruspini subforest contained in $\Nover_n$.
\item[\textup{(}iii\textup{)}] $\vdash_{\GL_4}\,\alpha \wedge \beta$, where
\begin{equation}\label{eq:formula4}
\alpha = \alpha_P \leftrightarrow \bigvee_{{l\in\Leaf(F_{\alpha_P})\cap\Leaf(\Rus_n)}}\psi_l\,,
\end{equation}
$$\beta = (\alpha_P\rightarrow\tau_n)\,.$$
\end{enumerate}
Moreover, for any Ruspini subforest $F$ contained in $\Nover_n$ there exists a
$2$-overlapping Ruspini partition $P'=\{f'_1,\dots,f'_n\}$, with $f'_i:[0,1]\to[0,1]$, such that $F(P') = F$.
\end{thm}
\begin{pf}
(\emph{i}) $\Rightarrow$ (\emph{ii}). By Theorem \ref{th:main}, $F(P)$ is a Ruspini subforest.
By Theorem \ref{th:main2}, $F(P)\subseteq\Nover_n$.

 (\emph{ii}) $\Rightarrow$ (\emph{iii}). By Theorem \ref{th:main}, the formula
$\alpha$
is a tautology of $\GL_\infty$, and thus a tautology of $\GL_4$. By Theorem \ref{th:main2}, the formula
$\beta$ is a tautology of $\GL_4$.
Therefore, the formula $\alpha \wedge \beta$ is
a tautology of $\GL_4$.

 (\emph{iii}) $\Rightarrow$ (\emph{i}). Since $\alpha \wedge \beta$ is a tautology
of  $\GL_4$, we have $\vdash_{\GL_4}\, \beta$. By Theorem \ref{th:main2}, $P$ is $2$-overlapping.
Moreover, by Lemma \ref{l:ginf to g4}, $\vdash_{\GL_4}\, \beta$ implies $\vdash\, \beta$, and therefore
\begin{equation}\label{eq:forestinclusion}
F_{\alpha_P}\subseteq \Nover_n\,.
\end{equation}
It remains to show that $P$ is a weak Ruspini partition.
The argument is analogous to that in (\emph{iii}) $\Rightarrow$ (\emph{i}) of Theorem \ref{th:main}.
Indeed, we note that $\GL_4$ proves the formula $\alpha \leftrightarrow \beta$ if and only if
$F_\alpha \cap \F_n^4 = F_\beta \cap \F_n^4$. Moreover, by (\ref{eq:forestinclusion}),
the element $k$ appearing in conditions (\textit{a}) and (\textit{b}) in the proof of Theorem \ref{th:main}
necessarily belongs to $\tau_n$, and thus to $\F_n^4$.

The last statement of the theorem is an immediate consequence of Theorems \ref{th:main} and \ref{th:main2}.
\end{pf}
The analogue of Corollary \ref{cor:enum} for
$2$-overlapping weak Ruspini partition is as follows.
\begin{cor}
\label{cor:enum2}
The number of classes of \god-equivalent $2$-overlapping weak Ruspini partitions
of $n$ elements is $2^{\frac{3n^2-n}{2}}-1$.
\end{cor}
\begin{pf}
A $2$-overlapping weak Ruspini partition $P$ is characterized,
up to \god{} equivalence, by the forest $F(P)$, and therefore
by a subset of leaves of $\Rus_n \cap \Nover_n$. We observe
that, by Definitions \ref{def:ruspini} and \ref{def:non-overlapping forest},
$\Rus_n \cap \Nover_n$ is the forest
obtained by removing from $\Nover_n$
the single tree having height $1$, and the leaves of all the trees having height $2$.
Thus, $\Rus_n \cap \Nover_n$ contains exactly $n \choose 1$
forests of height $1$, and $n \choose 2$
forests of height $3$. Since the trees of height $3$ have precisely $3$ leaves,
the total number of leaves of $\Rus_n \cap \Nover_n$ is
$${n \choose 1} + 3{n \choose 2}= \frac{3n^2-n}{2}\,.$$

Noting that, for every $2$-overlapping weak Ruspini partition $P$,
$F(P)\neq\emptyset$, the corollary follows.
\end{pf}

\bibliographystyle{elsart-num}
\bibliography{Ecs_bibliography}

\begin{thebibliography}{10}
\expandafter\ifx\csname url\endcsname\relax
  \def\url#1{\texttt{#1}}\fi
\expandafter\ifx\csname urlprefix\endcsname\relax\def\urlprefix{URL }\fi

\bibitem{ruspini}
E.~Ruspini, A new approach to clustering, Information and Control 15 (1969)
  22--32.

\bibitem{tnorms}
E.~P. Klement, R.~Mesiar, E.~Pap, Triangular norms, Vol.~8 of Trends in
  Logic---Studia Logica Library, Kluwer Academic Publishers, Dordrecht, 2000.

\bibitem{hajek}
P.~H{\'a}jek, Metamathematics of fuzzy logic, Vol.~4 of Trends in
  Logic---Studia Logica Library, Kluwer Academic Publishers, Dordrecht, 1998.

\bibitem{gottwald}
S.~Gottwald, A treatise on many-valued logics, Vol.~9 of Studies in Logic and
  Computation, Research Studies Press Ltd., Baldock, 2001.

\bibitem{yenlan}
J.~Yen, R.~Langari, {F}uzzy {L}ogic: {I}ntelligence, {C}ontrol, and
  {I}nformation, Prentice-Hall, Upper Saddle River, New Jersey, 1998.

\bibitem{ecsqaru07}
P.~Codara, O.~M. D'Antona, V.~Marra, Best approximation of {R}uspini
  {P}artitions in {G}{\"o}del logic, in: K.~Mellouli (Ed.), ECSQARU 2007, Vol.
  4724 of LNCS (LNAI), Springer, 2007, pp. 161--172.

\bibitem{bd}
R.~Balbes, P.~Dwinger, Distributive lattices, University of Missouri Press,
  Columbia, Mo., 1974.

\bibitem{horn}
A.~Horn, Free {$L$}-algebras, J. Symbolic Logic 34 (1969) 475--480.

\bibitem{APAL}
O.~M. D'Antona, V.~Marra, Computing coproducts of finitely presented {G}\"odel
  algebras, Ann. Pure Appl. Logic 142~(1-3) (2006) 202--211.

\bibitem{ieee}
P.~Codara, O.~M. D'Antona, V.~Marra, {P}ropositional {G}{\"o}del logic and
  {D}elannoy paths, IEEE International Fuzzy Systems Conference (FUZZ-IEEE)
  (2007) 1--5.

\bibitem{bcf}
M.~Baaz, A.~Ciabattoni, C.~G. Ferm{\"u}ller, Hypersequent calculi for {G}\"odel
  logics---a survey, J. Logic Comput. 13~(6) (2003) 835--861.

\bibitem{rourke}
C.~P. Rourke, B.~J. Sanderson, Introduction to piecewise-linear topology,
  Springer Study Edition, Springer-Verlag, Berlin, 1982, reprint.

\bibitem{baaz}
M.~Baaz, H.~Veith, Interpolation in fuzzy logic, Arch. Math. Logic 38~(7)
  (1999) 461--489.

\bibitem{aguzzoli}
S.~Aguzzoli, B.~Gerla, C.~Manara, Poset representation for {G}{\"o}del and
  {N}ilpotent {M}inimum logics, in: L.~Godo (Ed.), ECSQARU 2005, Vol. 3571 of
  LNCS (LNAI), Springer, 2005, pp. 662--674.

\bibitem{sloane}
N.~J.~A. Sloane, The on-line encyclopedia of integer sequences, Published
  electronically at http://www.research.att.com/�njas/sequences/ (2006).

\end{thebibliography}

\end{document}